\documentclass{svjour3}                     
\smartqed  
\usepackage{graphicx}
\usepackage{amsmath}
\journalname{Few-Body Systems (FB20)}
\begin{document}

\title{
No-Core Shell Model Analysis of Light Nuclei
\thanks{Computing support for this work came from the LLNL institutional Computing Grand Challenge program and the J\"ulich supercomputer Centre. Prepared in part by LLNL under Contract DE-AC52-07NA27344. Support from the U.\ S.\ DOE/SC/NP (Work Proposal No.\ SCW1158), the NSERC Grant No. 401945-2011, and from the U.\ S.\ Department of Energy Grant DE-FC02-07ER41457 is acknowledged. TRIUMF receives funding via a contribution through the Canadian National Research Council. This work is supported in part by the Deutsche Forschungsgemeinschaft through contract SFB 634 and by the Helmholtz International Center for FAIR within the framework of the LOEWE program launched by the State of Hesse.}
\thanks{Presented at the 20th International IUPAP Conference on Few-Body Problems in Physics, 20 - 25 August, 2012, Fukuoka, Japan}
}


\author{Sofia Quaglioni \and Petr Navr\'atil \and \\
              Guillaume Hupin \and Joachim Langhammer \and
              Carolina Romero-Redondo \and Robert Roth
}


\institute{S. Quaglioni \at
              Lawrence Livermore National Laboratory, P.O. Box 808, L-414, Livermore, CA 94551, USA \\
              Tel.: +1-925-4228152\\
              Fax: +1-925-4225940\\
              \email{quaglioni1@llnl.gov}           
           \and
           P. Navr\'atil \at
              TRIUMF, 4004 Wesbrook Mall, Vancouver, BC V6T 2A3, Canada;
              Lawrence Livermore National Laboratory, P.O. Box 808, L-414, Livermore, CA 94551, USA   
           \and
           G. Hupin \at
            Lawrence Livermore National Laboratory, P.O. Box 808, L-414, Livermore, CA 94551, USA 
            \and
            J. Langhammer \at
            Institut f\"{u}r Kernphysik, Technische Universit\"{a}t Darmstadt, 64289 Darmstadt, Germany
            \and
            C. Romero-Redondo  \at
            TRIUMF, 4004 Wesbrook Mall, Vancouver, BC V6T 2A3, Canada
            \and
            R. Roth \at
            Institut f\"{u}r Kernphysik, Technische Universit\"{a}t Darmstadt, 64289 Darmstadt, Germany
              }

\date{Received: date / Accepted: date}

\maketitle

\begin{abstract}
The fundamental description of both structural properties and reactions of light nuclei in terms of constituent protons and neutrons interacting through nucleon-nucleon and three-nucleon forces is a long-sought goal of nuclear theory. I will briefly present a promising technique, built upon the {\em ab initio} no-core shell model, which emerged recently as a candidate to reach such a goal: the no-core shell model/resonating-group method. This approach, capable of describing simultaneously both bound and scattering states in light nuclei, complements a microscopic cluster technique with the use of two-nucleon realistic interactions, and a microscopic and consistent description of the nucleon clusters. I will discuss applications to light nuclei binary scattering processes and fusion reactions that power stars and Earth based fusion facilities, such as the deuterium-$^3$He fusion, and outline the progress toward the inclusion of the three-nucleon force into the formalism and the treatment of three-body clusters.
\end{abstract}

\section{Introduction}
\label{intro}
Nuclei are aggregates of protons and neutrons interacting through forces arising from the underlying theory of quantum chromodynamics. 
Understanding how the strong force binds nucleons together in nuclei is fundamental to explain the very existence of the universe. Indeed, the mutual interactions between nucleons led to the formation of the lightest nuclei a few minutes after the Big Bang, and  the following nuclear processes, producing  heavier nuclei during stellar evolution and in violent events like supernovae, have been crucial in shaping the world we leave in. Therefore, one of the central goals of nuclear physics is to come to a basic understanding of the structure and dynamics of nuclei. The {\em ab initio} ({\em i.e.} from first principles) no-core shell model/resonating group method (NCSM/RGM)~\cite{PhysRevLett.101.092501,PhysRevC.79.044606} is a theoretical technique that attempts to achieve such a goal for light nuclei.  

\section{{\em Ab initio} NCSM/RGM}
\label{formalism}
In the {\em ab initio} NCSM/RGM approach the many-body wave function, 
\begin{eqnarray}
|\Psi^{J^\pi T}\rangle &=& \sum_{\nu} \int dr r^2 \, \hat{\mathcal A}_{\nu}|\Phi^{J^\pi T}_{\nu r}\rangle  \frac{[{\cal N}^{-1/2}\chi]^{J^\pi T}_\nu(r)}{r}
\, , \label{trial}
\end{eqnarray}
is expanded over a set of  translational-invariant cluster basis states describing two nuclei (a target and a projectile composed of $A-a$ and $a \le A$ nucleons, respectively) whose centers of mass are separated by the relative coordinate $\vec r_{A-a,a}$ and that are traveling in a $^{2s}\ell_J$ wave or relative motion (with $s$ the channel spin, $\ell$ the relative momentum, and $J$ the total angular momentum of the system):
\begin{equation}
|\Phi^{J^\pi T}_{\nu r}\rangle = \Big [ \big ( \left|A{-}a\, \alpha_1 I_1^{\,\pi_1} T_1\right\rangle \left |a\,\alpha_2 I_2^{\,\pi_2} T_2\right\rangle\big ) ^{(s T)}
Y_{\ell}\left(\hat r_{A-a,a}\right)\Big ]^{(J^\pi T)}\frac{\delta(r-r_{A-a,a})}{rr_{A-a,a}}\,.\label{basis}
\end{equation}
Here, the antisymmetric wave functions $\left|A{-}a\, \alpha_1 I_1^{\,\pi_1} T_1\right\rangle$ and $\left |a\,\alpha_2 I_2^{\,\pi_2} T_2\right\rangle$ are eigenstates of the $(A-a)$- and $a$-nucleon intrinsic Hamiltonians, respectively, as obtained within the NCSM approach~\cite{PhysRevLett.84.5728} and are characterized by the spin-parity, isospin and energy labels $I_i^{\pi_i},T_i$, and $\alpha_i$, respectively, where $i=1,2$. Additional quantum numbers labeling these RGM-inspired continuous basis states are  parity $\pi=\pi_1\pi_2(-1)^{\ell}$ and total isospin $T$. In our notation, all these quantum numbers are grouped into a cumulative index $\nu=\{A{-}a\,\alpha_1I_1^{\,\pi_1} T_1;\, a\, \alpha_2 I_2^{\,\pi_2} T_2;\, s\ell\}$. The Pauli principle is enforced by introducing the appropriate inter-cluster antisymmetrizer, schematically
\begin{equation}
\hat{\mathcal A}_{\nu}=\sqrt{\frac{(A{-}a)!a!}{A!}}\left( 1+\sum_{P\neq id}(-)^pP\right)\,,
\label{antisymmetrizer}
\end{equation}   
where the sum runs over all possible permutations of nucleons $P$ different from the identical one 
that can be carried out between the two different clusters, 
and $p$ is the number of interchanges characterizing them. 
Finally, the continuous linear variational amplitudes $\chi^{J^\pi T}_\nu(r)$  are determined by solving the orthogonalized RGM equations: 
\begin{equation}
{\sum_{\nu^\prime}\int dr^\prime r^{\prime\,2}} [{\mathcal N}^{-\frac12}{\mathcal H}\,{\mathcal N}^{-\frac12}]^{J^\pi T}_{\nu\nu^\prime\,}(r,r^\prime)\frac{\chi^{J^\pi T}_{\nu^\prime} (r^\prime)}{r^\prime} = E\,\frac{\chi^{J^\pi T}_{\nu} (r)}{r}  \label{RGMeq}\,,
\end{equation}
where ${\mathcal N}^{J^\pi T}_{\nu\nu^\prime}(r, r^\prime)$ and ${\mathcal H}^{J^\pi T}_{\nu\nu^\prime}(r, r^\prime)$, commonly referred to as integration kernels, are respectively the overlap (or norm) and Hamiltonian matrix elements over the antisymmetrized basis~(\ref{basis}), {\em i.e.}:  
\begin{align}
{\mathcal N}^{J^\pi T}_{\nu^\prime\nu}(r^\prime, r) = \left\langle\Phi^{J^\pi T}_{\nu^\prime r^\prime}\right|\hat{\mathcal A}_{\nu^\prime}\hat{\mathcal A}_{\nu}\left|\Phi^{J^\pi T}_{\nu r}\right\rangle, 
&\;
{\mathcal H}^{J^\pi T}_{\nu^\prime\nu}(r^\prime, r) = \left\langle\Phi^{J^\pi T}_{\nu^\prime r^\prime}\right|\hat{\mathcal A}_{\nu^\prime}H\hat{\mathcal A}_{\nu}\left|\Phi^{J^\pi T}_{\nu r}\right\rangle.
\label{NH-kernel}
\end{align}
Here,  $H$ is the microscopic $A-$nucleon Hamiltonian and $E$ is the total energy in the center of mass (c.m.)\ frame.
For a detailed explanation of how norm and Hamiltonian kernels are obtained from the underlying nuclear interaction and the NCSM eigenvectors of  target and projectile we refer the interested reader to Refs.~\cite{PhysRevC.79.044606} and~\cite{PhysRevC.83.044609}. 
 
\section{Applications}
Applications of the NCSM/RGM approach to the description of nucleon- and deuteron-nucleus
type of collisions based on two-nucleon ($NN$) realistic interactions have already led to
very promising results~\cite{PhysRevLett.101.092501,PhysRevC.79.044606,PhysRevC.82.034609,PhysRevC.83.044609,Navratil2011379,PhysRevLett.108.042503}. In most instances, we employed similarity-renormalization-group (SRG)~\cite{PhysRevC.75.061001,PhysRevC.77.064003} evolved potentials, and in particular, those obtained from the chiral N$^3$LO~\cite{N3LO} $NN$ interaction. Here we briefly review two of such applications, the calculation of the $^7$Be$(p,\gamma)^8$B radiative capture~\cite{Navratil2011379}, and  the study of the $^3$H$(d,n)^4$He and $^3$He$(d,p)^4$He fusion reactions~\cite{PhysRevLett.108.042503}.

\subsection{The $^7$Be$(p,\gamma)^8$B radiative capture}
Recently, we have performed the first {\em ab initio} many-body calculation of the $^7$Be$(p,\gamma)^8$B radiative capture~\cite{Navratil2011379}, the final step in the nucleosynthetic chain leading to $^8$B and one of the main inputs of the Standard Solar Model. This calculation was carried out in a model space spanned by $p$-$^7$Be channel states including the five lowest eigenstates of $^7$Be (the $\tfrac32^-$ ground and the $\tfrac12^-$,$\tfrac72^-$, and first and second $\tfrac52^-$ excited states) in an $N_{\rm max}=10$ NCSM basis, and employed the SRG-N$^3$LO $NN$ interaction with $\Lambda=1.86$ fm$^{-1}$, where $\Lambda$ denotes the SRG evolution parameter~\cite{PhysRevC.75.061001}. 
We first solved Eq.~(\ref{RGMeq}) with bound-state boundary conditions to find the bound state of $^8$B, and then with scattering boundary conditions to find the $p$-$^7$Be scattering wave functions. Former and latter wave functions were later used to calculate the capture cross section, which, at solar energies, is dominated by non-resonant $E1$ transitions from $p$-$^7$Be $S$- and $D$-waves into the weakly-bound ground state of $^8$B.  All stages of the calculation were based on the same HO frequency of $\hbar\Omega=18$ MeV, which minimizes the g.s.\ energy of $^7$Be. 

At $N_{\rm max}=10$, the largest model space achievable for the present calculation within the full NCSM basis,  the $^7$Be g.s. energy is very close to convergence as indicated by a fairly flat frequency dependence in the range $16\le\hbar\Omega\le20$ MeV, and the vicinity to the $N_{\rm max}=12$ result obtained within the importance-truncated NCSM~\cite{PhysRevLett.99.092501,PhysRevC.79.064324}.  With the chosen value $\Lambda=1.86$  fm$^{-1}$ for the SRG evolution of the N$^3$LO $NN$ interaction, we obtain a single $2^+$ bound state for $^8$B with a separation energy of 136 keV, which is quite close to the observed one (137 keV). This is very important for the description of the low-energy behavior of the  $^7$Be$(p,\gamma)^8$B astrophysical S-factor, known as $S_{17}$. While for a complete {\em ab initio}  calculation one should include also the three-nucleon ($NNN$) interaction induced by the SRG evolution of the $NN$ potential as well as the SRG-evolved attractive initial chiral $NNN$ force, we note that in the $\Lambda$-range $\sim1.8$-$2.1$ fm$^{-1}$, and, in very light nuclei, the former interaction is repulsive and the two contributions cancel each other to a good extent~\cite{PhysRevLett.103.082501,PhysRevC.83.034301}.   
\begin{figure*}[t]
\begin{minipage}{13.5pc}
\includegraphics[width=13.5pc]{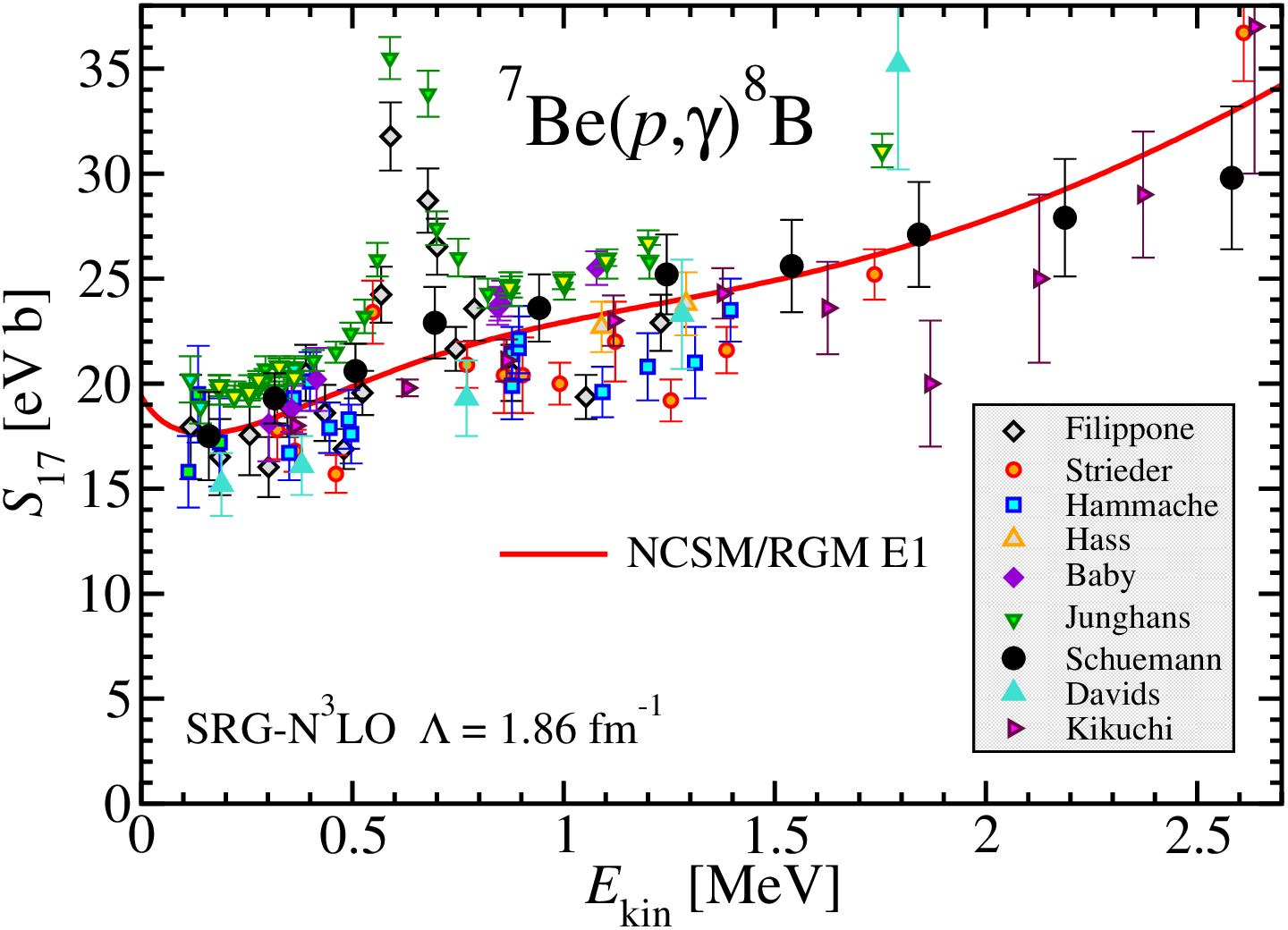}
\caption{\label{fig:p7Be1}Calculated $^7$Be$(p,\gamma)^8$B S-factor as a function of the energy in the center of mass compared to data. Only $E1$ transition were considered in the calculation.}
\end{minipage}\hspace{2pc}%
\begin{minipage}{13.5pc}
\includegraphics[width=13.5pc]{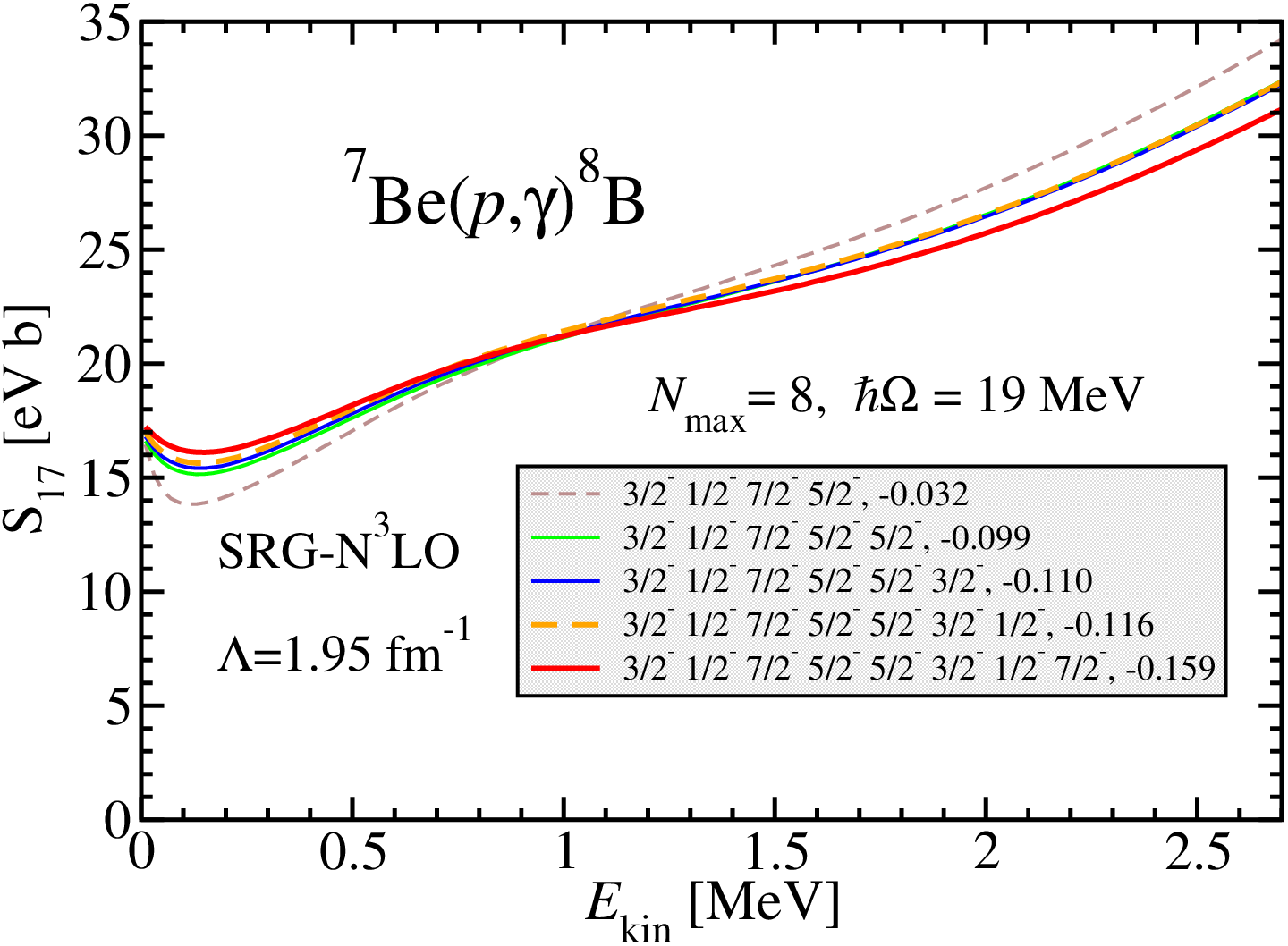}
\caption{\label{fig:p7Be2}Convergence of the $^7$Be$(p,\gamma)^8$B S-factor as a function of the number of $^7$Be eigenstates included in the calculation (shown in the legend together with the corresponding separation energy). }
\end{minipage} 
\end{figure*}

Figure 1 compares the resulting $S_{17}$ astrophysical factor with several experimental data sets. Energy dependence and absolute magnitude follow closely the trend of the indirect Coulomb breakup measurements of Sh\"umann {et al}.~\cite{Schuemann1}, 
while somewhat underestimating the direct data of Junghans {\em et al}.~\cite{Junghans}. The resonance due to the $M1$ capture, particularly evident in these and Filippone's data and missing in our results, does not contribute to a theoretical calculation outside of the narrow $^8$B $1^+$ resonance and is negligible at astrophysical energies~\cite{Adelberger1}. 
The $M1$ operator, for which any dependence upon two-body currents needs to be included explicitly, poses more uncertainties than the Siegert's $E1$ operator. In addition, the treatment of this operator within the NCSM/RGM approach is slightly complicated by the contributions coming from the core ($^7$Be) part of the wave function. Nevertheless, we plan to calculate its contribution in the future. 

Our calculated $S_{17}(0)=19.4(7)$ MeV b is on the lower side, but consistent with the latest evaluation $20.8\pm0.7$(expt)$\pm1.4$(theory)~\cite{Adelberger1}. The  0.7 eV b uncertainty was estimated by studying the dependence of the S-factor on the harmonic oscillator (HO) basis size $N_{\rm max}$ as well as the influence of higher-energy excited states of the $^7$Be target.  More precisely, we performed calculations up to $N_{\rm max}=12$ within the importance-truncation NCSM scheme~\cite{PhysRevLett.99.092501,PhysRevC.79.064324} including (due to computational limitations) only the first three eigenstates of $^7$Be. The $N_{\rm max}=10$ and $12$ S-factors are very close. In addition, the convergence in the number of $^7$Be states was explored by means of calculations including up to 8 $^7$Be eigenstates in a $N_{\rm max}=8$ basis (larger $N_{\rm max}$ values are currently out of reach with more then five $^7$Be states).  This last set of calculations is presented in Fig.~\ref{fig:p7Be2}, from which it appears that, apart from the two $\tfrac52^-$ states, the only other state to have a significant impact on the $S_{17}$ is the second $\frac72^-$, the inclusion of which affects the separation energy and contributes somewhat to the flattening of the $S$-factor around $1.5$ MeV.  For these last set of calculations we used SRG-N$^3$LO interactions obtained with different $\Lambda$  values with the intent to match closely the experimental separation energy in each of the largest model spaces. Based on this analysis, we conclude that the use of an $N_{\rm max}=10$ HO model space is justified and the restriction to five $^7$Be eigenstates is quite reasonable. 

\subsection{The $^3$H$(d,n)^4$He and $^3$He$(d,p)^4$He fusion reactions}
\begin{figure*}[t]
\begin{minipage}{13.5pc}
\includegraphics[width=13.5pc]{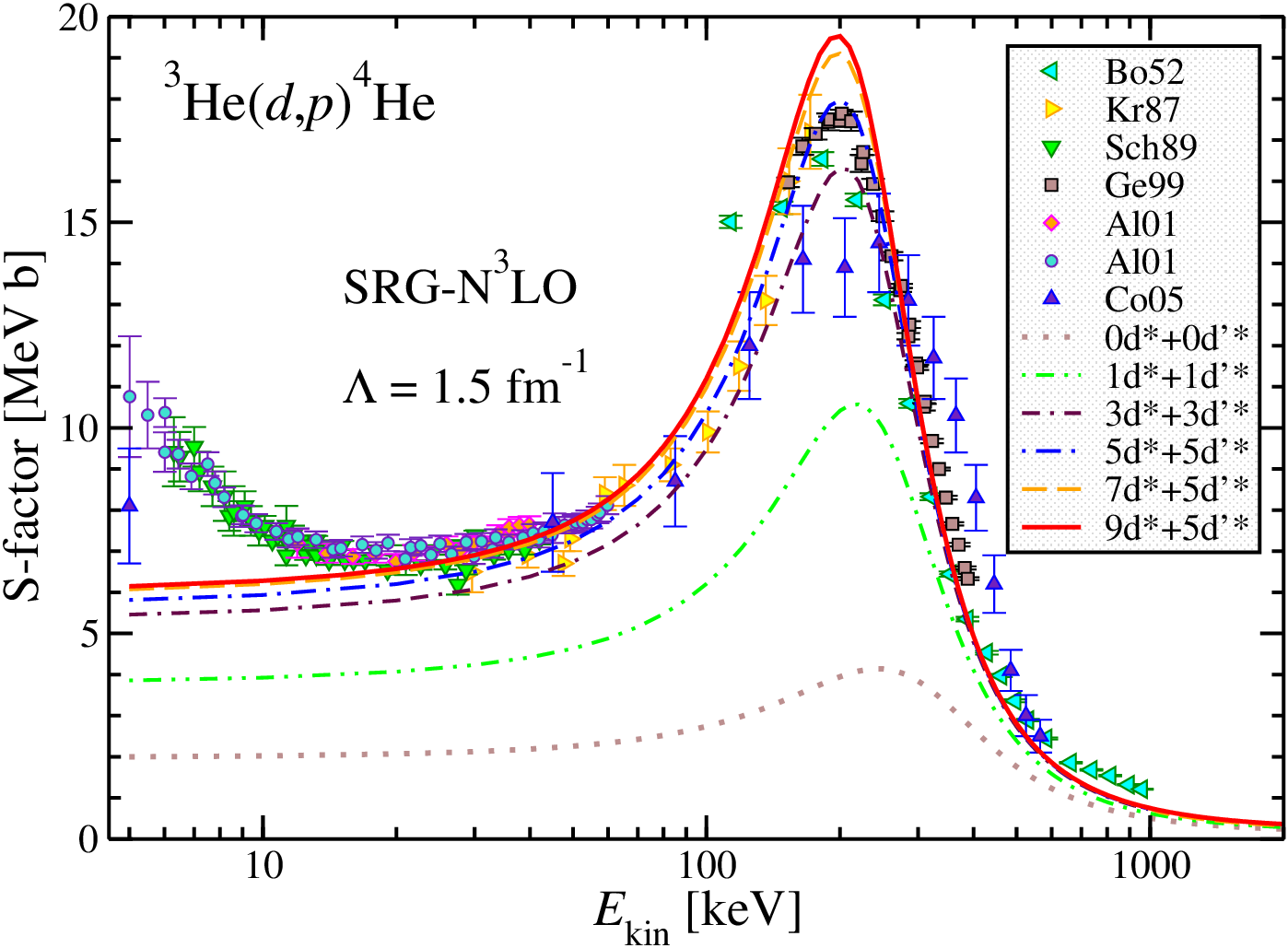}
\caption{\label{fig:d3He}Calculated S-factor of the $^3$He$(d,p)^4$He reaction compared to experimental data. Convergence with the number of deuterium pseudostates in the $^3S_1$-$^3D_1$ ($d^*$) and $^3D_2$ ($d^{\prime *}$) channels. }
\end{minipage}\hspace{2pc}%
\begin{minipage}{13.5pc}
\includegraphics[width=13.5pc]{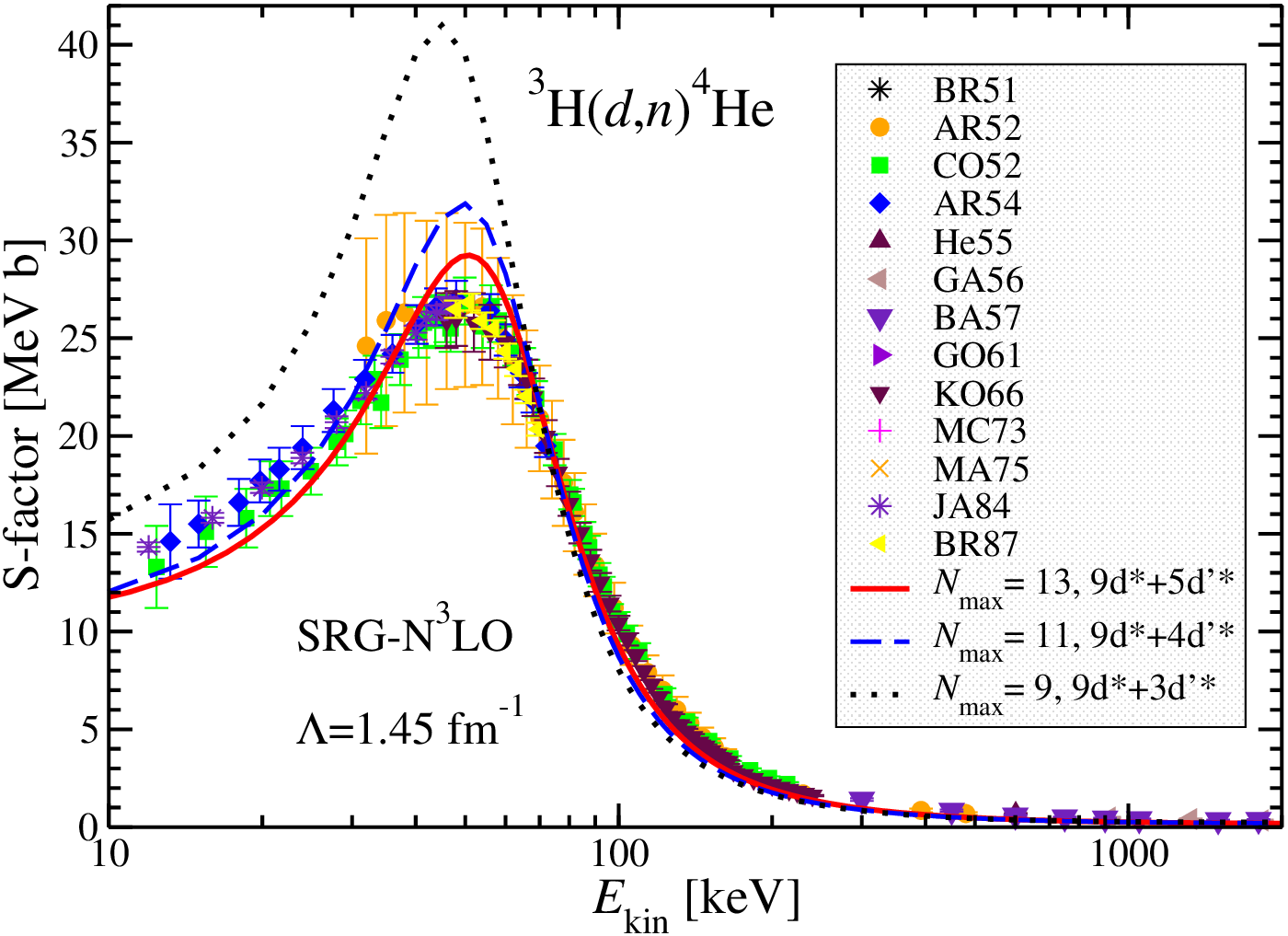}
\caption{\label{fig:d3H}Calculated $^3$H$(d,n)^4$He S-factor compared to experimental data. Convergence with $N_{\rm max}$ obtained for the SRG-N$^3$LO $NN$ potential with $\Lambda=1.45$ fm$^{-1}$ at $\hbar\Omega=14$ MeV.}
\end{minipage} 
\end{figure*}
In the following we present the first {\em ab initio} many-body calculations of $^3$H$(d,n)^4$He and $^3$He$(d,p)^4$He fusion reactions~\cite{PhysRevLett.108.042503} starting from the SRG-N$^3$LO $NN$ interaction with $\Lambda=1.5$ fm$^{-1}$, for which we reproduce the experimental $Q$-value of both reactions within $1\%$. These reactions have important implications first and foremost for fusion energy generation, but also for nuclear astrophysics, and atomic physics. Indeed, the deuterium-tritium fusion is the easiest reaction to achieve on earth and is pursued by research facilities directed at reaching fusion power. Both $^3$H$(d,n)^4$He and $^3$He$(d,p)^4$He  affect the predictions of Big Bang nucleosynthesis for light-nucleus abundances. In addition, the deuterium-$^3$He fusion is also an object of interest for atomic physics, due to the substantial electron-screening effects presented by this reaction. 

The model spaces adopted are characterized by HO model basis sizes up to $N_{\rm max}=13$ with a frequency of $\hbar\Omega=14$ MeV and channel bases including $n$-$^4$He ($p$-$^4$He), $d$-$^3$H ($d$-$^3$He), $d^*$-$^3$H ($d^*$-$^3$He) and $d^{\prime*}$-$^3$H ($d^{\prime*}$-$^3$He) binary cluster states.  Here, $d^*$ and $d^{\prime*}$ denote $^3S_1$-$^3D_1$ and $^3D_2$ deuterium excited pseudostates, respectively, and the $^3$H ($^3$He) and $^4$He nuclei are in their ground state. 

The results obtained for the $^3$He$(d,p)^4$He S-factor are presented in Figure~\ref{fig:d3He}. The deuteron deformation and its virtual breakup, approximated by means of $d$ pseudostates, play a crucial role in reproducing the observed magnitude of the S-factor. Convergence is reached for $9d^*+5d^{\prime*}$. The typical dependence upon the HO basis sizes adopted is illustrated by the $^3$H$(d,n)^4$He results of Fig.~\ref{fig:d3H}. The convergence is satisfactory and we expect that an $N_{\rm max}=15$ calculation, which is currently out of reach, would not yield significantly different results. While the experimental position of the $^3$He$(d,p)^4$He S-factor is reproduced within few tens of keV and we find an overall fair agreement with experiment (if we exclude the region at very low energy, where the accelerator data are enhanced by laboratory electron screening), the $^3$H$(d,n)^4$He S-factor is not described as well with  $\Lambda=1.5$ fm$^{-1}$. 
Due to its very low activation energy, the $^3$H$(d,n)^4$He  S-factor, particularly the position and height of its peak, is extremely sensitive to higher-order effects in the nuclear interaction, such as the $NNN$ force (not yet included in the calculation) and missing isospin-breaking effects in the integration kernels (which are obtained in the isospin formalism).  With a very small change in the value of the SRG evolution parameter we can compensate for these missing higher-order effects in the interaction and reproduce the position of the  $^3$H$(d,n)^4$He S-factor.  This led to the theoretical S-factor of Fig.~\ref{fig:d3H} (obtained for $\Lambda=1.45$ fm$^{-1}$), that is in overall better agreement with data, although it presents a slightly narrower and somewhat overestimated peak. This calculation would suggest that some electron-screening enhancement could also be present in the $^3$H$(d,n)^4$He measured S-factor below ~10 keV c.m.\ energy. However, these results cannot be considered conclusive until more accurate calculations using a complete nuclear interaction (that includes the $NNN$ force) are performed. Work in this direction is under way.

\section{Recent developments}
Here we outline some of our more recent efforts in the development of the NCSM/RGM approach, namely the progress toward the inclusion of the three-nucleon force into the formalism, and the treatment of three-body clusters and their dynamics.

\subsection{Scattering and three-nucleon force}
\begin{figure*}[t]
\begin{minipage}{13.5pc}
\includegraphics[width=13.5pc]{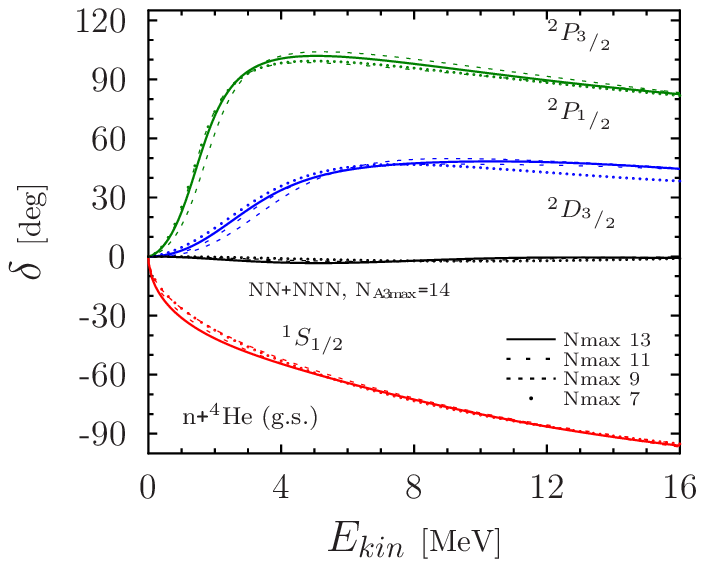}
\caption{\label{fig:n4He-1} Convergence with respect to the HO basis size $N_{\rm max}=11$ at $\hbar\Omega=20$ MeV of the $n+^4$He(g.s.) phase shifts obtained for the SRG-(N$^3$LO NN + N$^2$LO NNN) interaction with $\Lambda=2.0$ fm$^{-1}$. The label $N_{\rm A3max}=14$ refers to the HO size of the NNN matrix elements.}
\end{minipage}\hspace{2pc}%
\begin{minipage}{13.5pc}
\includegraphics[width=13.5pc]{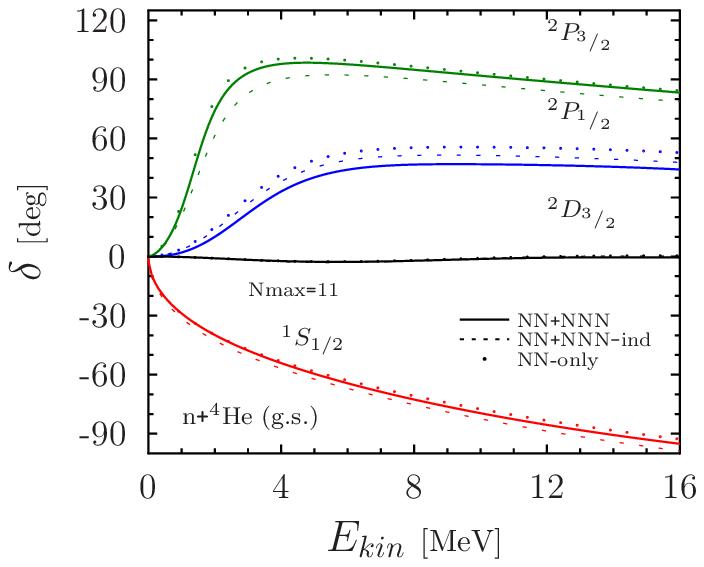}
\caption{\label{fig:n4He-2} Calculated $n+^4$He(g.s.) phase shifts for SRG-N$^3$LO NN-only (dots), NN+NNN-induced (dashed line) and SRG-(N$^3$LO NN + N$^2$LO NNN) interactions (solid line) with $\Lambda=2.0$ fm$^{-1}$ obtained within the HO basis size $N_{\rm max}=11$ and frequency $\hbar\Omega=20$ MeV.  See also the caption of Fig.~\ref{fig:n4He-1}.}
\end{minipage} 
\end{figure*}
In past applications for light-ion reactions, we omitted the $NNN$ interaction induced by the SRG renormalization of  the $NN$ potential as well as the initial chiral $NNN$ force. By neglecting induced forces, we introduced a dependence on the SRG parameter $\Lambda$, which was then chosen so that the particle separation energies were well reproduced. For low-energy thermonuclear reactions, this is a dominant effect, and overall such a procedure led to (never obtained before) realistic results. However, 
a truly accurate {\em ab initio} description demands the inclusion of both induced and chiral $NNN$ interactions. 

While the inclusion of the $NNN$ force into the NCSM/RGM formalism is conceptually straightforward, in practice it poses major challenges having to deal with: 1) the large number of $NNN$ matrix elements, which makes it essential to work within the JT-coupled scheme; and 2) the appearance of kernels depending on the three-body densities of the target already for nucleon-nucleus processes, which demands new efficient computing strategies for applications with $p$-shell targets to be possible. 
Figures~\ref{fig:n4He-1} and \ref{fig:n4He-2} present initial results for $^4$He$(n,n)^4$He scattering phase shifts with inclusion of the $NNN$ force. The use of SRG-evolved interactions facilitates the convergence of the calculation within $N_{\rm max}\sim11$. At the same time, Fig.~\ref{fig:n4He-2} highlights the influence of induced and initial components of the $NNN$ force on the resonant phase shifts. The largest splitting between $^2P_{3/2}$ and $^2P_{1/2}$ is found when both induced and chiral $NNN$ forces are included (NN+NNN curve). 
It should be noticed that these results are still preliminary, as not all relevant excitations of the $^4$He nucleus have been taken into account yet. In particular, 
the resonances are sensitive to the inclusion of the first six excited states of the $^4$He~\cite{PhysRevC.79.044606} (here only the g.s. is included). We will complete this study in the coming months. 

We have also obtained first results for the $d+^4$He scattering phase shifts and the ground state of $^6$Li including both SRG-induced and chiral $NNN$ forces.  The model spaces adopted so far contain only the g.s. of the $^4$He and $d$ nuclei within a $N_{\rm max} = 8$ HO basis size. While calculations for larger $N_{\rm max}$ values and including pseudo-excited states of the deuteron (fundamental to model the deformation and virtual breakup of this nucleus) are under way, these preliminary results are already very promising.  In particular, the inclusion of the $NNN$ force produces a change in position and splitting of the $^3D_1$ and $^3D_2$ scattering phase shifts, which were not well described in our former calculation with only the $NN$ part of the SRG interaction. The $NNN$ force has also the effect of increasing the binding energy of the $^6$Li nucleus.

\subsection{Three-cluster dynamics}
A proper description of Borromean halo nuclei and three-body breakup reactions (but also virtual breakup effects) within the NCSM/RGM approach requires the inclusion of three-cluster channel states and the treatment of the three-body dynamics.  
\begin{table}[!h]
\caption{Ground-state energies of the $^{4,6}$He nuclei in MeV. Both NCSM/RGM and NCSM calculations were performed with the SRG-N$^3$LO $NN$ potential with $\Lambda=1.5$ fm$^{-1}$, and $\hbar\Omega=16$ MeV HO frequency. Extrapolations were performed with an exponential fit.}
\label{tab:1}       
\begin{tabular}{llll}
\hline\noalign{\smallskip}
Approach & &$E_{\rm g.s.}(^4$He$)$ & $E_{\rm g.s.}(^6$He$)$ \\
\noalign{\smallskip}\hline\noalign{\smallskip}
NCSM/RGM &($N_{\rm max}=12$) & $-28.22$ MeV & $-28.72$ MeV \\
NCSM &($N_{\rm max}=12$) & $-28.22$ MeV & $-29.75$ MeV \\
NCSM &(extrapolated) & $-28.23$ MeV&  $-29.80$ MeV\\
\noalign{\smallskip}\hline
\end{tabular}
\end{table}
At present we have completed the development of the formalism for the treatment of three-cluster systems formed by two separate nucleons in relative motion with respect to a nucleus of mass number $A-2$.  
Preliminary results for the g.s.\ energy of  $^6$He within a $^4$He(g.s.)$+ n + n$ cluster basis and an $N_{\rm max}=12$, $\hbar\Omega=16$ MeV HO model space, are compared to NCSM calculations in  Table~\ref{tab:1}. The interaction adopted is the SRG-N$^3$LO $NN$ potential with $\Lambda=1.5$ fm$^{-1}$. With such a low value of $\Lambda$, at $N_{\rm max}\sim12$ the binding energy calculations are close to convergence in both NCSM/RGM and NCSM approaches. The $\sim 1$ MeV difference observed is due to the excitations of the $^4$He core, included only in the NCSM at present. Contrary to the NCSM, in the NCSM/RGM the $^4$He(g.s.)$+n+n$ wave functions present the appropriate asymptotic behavior. 
This will be essential in describing $^6$He excited states in the continuum, such as, {\em e.g.}\ the $1^-$ soft dipole resonance. Work towards the solution of the three-cluster NCSM/RGM equations with continuum boundary conditions is under way.

\section{Conclusions}
We presented an outline of the NCSM/RGM, an {\em ab initio} many-body approach capable of describing simultaneously both bound and scattering states in light nuclei, by complementing the RGM with the use of realistic interactions, and a microscopic and consistent description of the nucleon clusters, obtained via the {\em ab initio} NCSM. We discussed applications to fusion reactions that power stars and Earth based fusion facilities, such as the $^7$Be$(p,\gamma)^8$B radiative capture, and the $^3$H$(d,n)^4$He and $^3$He$(d,p)^4$He fusion reactions. Finally, we outlined the progress toward the inclusion of the $NNN$ force into the formalism and the treatment of three-cluster dynamics, and presented an initial assessment of $NNN$-force effects on $^4$He$(n,n)^4$He scattering, as well as preliminary results for the g.s.\ energy of $^6$He computed within a $^4$He(g.s.)$+n+n$ NCSM/RGM three-cluster basis. Since the publication of the first results~\cite{PhysRevLett.101.092501,PhysRevC.79.044606,PhysRevC.82.034609}, 
obtained for nucleon-nucleus collisions, the NSCM/RGM has grown into a powerful approach for the description of binary reactions starting from realistic $NN$ forces. A truly accurate {\em ab initio} description of light-ion fusion reactions and light exotic nuclei that encompasses the full $NNN$ force and the three-cluster dynamics is now within reach. 



\end{document}